\documentclass[a4paper,10pt]{article}
\usepackage{amsmath}
\usepackage{amsfonts}
\usepackage{amssymb}

\title{Third Spectroscopy with a hint of superstrings}
\author{A. Rivero \thanks{Institute for Biocomputation and
Physics of Complex Systems, University of Zaragoza} \thanks{\texttt{al.rivero@gmail.com}, \texttt{arivero@unizar.es}}}

\begin{document}

\maketitle

\begin{abstract}
Some new regularities found in the pseudoscalar meson octet are reported. They invite to reconsider models where elementary fermions and composite QCD open strings can be grouped in common supermultiplets.
\end{abstract}

\section{Introduction}

It was Weisskopf who coined the term ``\textit{third spectroscopy}'' to refer to the subnuclear energy range and to stress the empiricall parallel with atomic and nuclear spectroscopies. While it is true that many of the levels of the hadron resonances can be understood with the systematics learned from first and second spectroscopies, it is also a disturbing point that the most fundamental quantities are free parameters. It is particularly intriguing that the general organization of the QCD sector is patterned in the same four sectors than the GWS sector. Legend says that the discovery of the muon was during some months mistaken with the prediction of the pion, and it is still unknown if there is a deeper reason for the proximity of the masses of both particles, an elementary and a composite.

For three years, the internet thread

\textbf{http://www.physicsforums.com/showthread.php?t=46055}

has been acting as a clearinghouse for amateur spectroscopists to inform about patterns they come to observe in the mass and widths of the cataloged particles. Of course such investigation is always menaced by the birthday paradox and other delusions, and it is fortunate that the contributors try to refrain themselves to relationships with high precision, usually greater than three digits. 

\section{Leptons, the octet and other histories}

With the breaking of flavour SU(3) to SU(2) isospin plus U(1) hypercharge, the pseudoscalar octet has three mass parameters linked via a Gell-Mann Okubo equation

$$
m_{\eta_8}^2 = \frac 13 (4 m_K^2-m_\pi^2)
$$

Using the measured values\footnote{Along this paper, when not said otherwise, values are quoted in \textit{MeV}. All the experimental values are taken from the compilation \cite{pdg}} $m_\pi=134.9766 \pm 0.0006$, for the neutral pion, and $m_K=497.648 \pm 0.022$, we get $m_{\eta_8}=569.326 \pm 0.026$. 
The mixing of this quantity with the singlet in order to produce $\eta, \eta'$ is the $U(1)$ problem, solved in part with 't Hooft's instantons, in part with Witten-Veneziano formula, and in part still a open question, which we do not address here.

We use the mass values where the charged leptons lay. Namely
$$
m_e=0.51099892 \pm 0.00000004; m_\mu=105.658369 \pm 0.000009;
m_\tau=1776.90 \pm 0.20  
$$

Note we do not say we use ``the values of the charged leptons''. The idea in the second part of this paper is that these are the mass values of whole supermultiplets, before further interaction acts in the meson side.

Now the newest remarks we want to report are the following ones:

\begin{eqnarray*}
  \sqrt{m_e}\sqrt{m_\tau-m_\mu}&=& 29.2233 \pm 0.0017 \\
  m_\pi-m_\mu&=&  29.3182 \pm 0.0006 \\
  \sqrt{m_\tau}\sqrt{m_\mu-m_e}&=&  432.246 \pm 0.024 \\
  \sqrt{m_\mu}\sqrt{m_\tau-m_e}&=&  433.232 \pm 0.024 \\
  m_{\eta_8}-m_\pi&=& 434.349 \pm 0.026 
\end{eqnarray*} 

The agreement between first and second result is over 99.6 \%, the agreement between fourth and fifth is over 99.7 \%.

The first two lines of the above array were noticed by T.A. Mir (aka Taarik) during a research on mass multiplicities \cite{Shah:2007gw} in the way of McGregor \cite{Mac Gregor:1980jua}. Then the last three ones were suggested by the author of this note. While Taarik's formula admits two different $S_3$ permutations, the implied values are very similar.

Early in this year, Hans de Vries had noticed than the breaking of isospin symmetry also has a simple relationship with the mass of the muon, namely,

$$
\left\vert {m_{\pi^+} \over m_\pi} -1 \right\vert^2 = { m_\mu \over m_Z}
$$

With $m_{\pi^+}=139.57018 \pm 0.00035 $ and $m_Z=91.1876 ± 0.0021 \mbox{GeV}$) 
the LHS is $  0.00115821   \pm .00000049$
and the RHS is  $0.00115869 \pm .00000003$. The central values agree within 99.95\%, the discrepancy being of the order of experimental error.

Perhaps this relationship was not noticed previously in the literature because the usual calculation (eg \cite{Witten:1983ut}) of isospin breaking is a additive perturbation, not a multiplicative adjust. A very similar formula was also noticed by HdV in the difference neutron-proton, pivoting on $m_{W^+}$. This is related to previously reported findings \cite{deVries:2005xi}.

Similarly, it was not obvious to spot the first set of relationships because the mass differences of the octet are usually worked out as squares $m_a^2 - m_b^2$. Also the use of square roots of the lepton mass values is a bit of surprise, but this use already happens in the famous relationship of Koide \cite{Koide:1983qe}, which other authors (eg CarlBrannen in our thread) have linked to a tribimaximal matrix of lepton mixing. Following a suggestion of R. Foot\cite{Foot:1994yn}, Koide's relationship can be written as an angle against the symmetric, diagonal, vector
$$
 (1,1,1) \measuredangle  (\sqrt{m_e},\sqrt{m_\mu},\sqrt{m_\tau}) = 45^\circ
$$

We have been recently aware of some new works \cite{masusy,ksusy, ksusy2} that rely on superpotentials to produce Koide equation. The original derivation was based in a preon model, and then it is interesting to link the formula to composites, even if we observe it in elementary objects.

Another point to remember is that, due to the failure of perturbation theory for QCD at low energies, the above quantities in the octet can only be calculated --while we wait for finer lattice results-- via chiral perturbation theory, a setup where the unknown information is swept under the carpet of decay constants. There is another situation where various results of chiral perturbation theory are intriguingly related between and with the parameters of electroweak theory: all the known complete electromagnetic decay widths of neutral particles, with the exception of the Upsilon meson, coincide when they are scaled with the cube of the mass of the decaying particle. And, they coincide with the total decay width of $Z^0$, if scaled with the same rule. This situation was also reported previously. The next table shows the initial observation, where $\eta$ and $\Sigma^0$ have also non-EM decay paths. See Fig. 2 in \cite{Rivero:2006rb} for more particles and full details.

\begin{center}
\begin{tabular}{r|llllll}
 & $\pi^0$ & $\eta$ & $\omega$ & $\Sigma^0$ & $J/\Psi$ & $Z^0$ \\ 
$(\Gamma_{\mbox{\begin{scriptsize}TOT.\end{scriptsize}}}/M^3)^{-1/2}$, (GeV) & 561 & 357 & - & 138 & 571 & 551 \\
$(\Gamma_{\mbox{\begin{scriptsize}EM.\end{scriptsize}}}/M^3)^{-1/2}$, (GeV)&         &   548 &591     &            &          &      
\end{tabular}
\end{center}

Note that the current uncertainty in the half life of the neutral pion allows to ``predict'' it via this naive scaling from $Z^0$, the result lying well inside the 1-sigma error bar. 

\section{Boson/Fermion d.o.f. and the string}

The features observed in the previous section involve relationships between composite bosons and elementary fermions, which is quite surprising. Now, all the composite bosons are QCD mesons, so it is possible to pursue some model in terms of a QCD SU(3) string. And, given the involvement of fermions, a supersymmetric model is desirable.

A first test to ask if superstring models are feasible is to count the degrees of freedom in each side.  The ``abstruse'' identity of Jacobi can not be fulfilled in any case, because in the elementary fermion side there can not be particles with spin 3/2 or greater. So we are only interested in matching spin 0 bosons against spin 1/2 fermions. And even to do this, we must take into account that not every quark terminates a string. In the standard model, the top quark does not bind into mesons.

In general, we can consider that our ``oriented string'' sector must be built from $n$ generations of quarks and leptons where only $r$ quarks of the D type and $s$ quarks of the U type bind into mesons. Furthermore, this binding is classified by a $SU(N), N=r+s$ , flavour group  and $N \otimes N^*$ decomposes into $(N^2-1) \oplus \mathbf{1}$, leaving out one degree of freedom from the neutral mesons \footnote{in some formalisms it is suggested that this extra degree is accounted by the closed string}.

So the coincidence of degrees of freedom for fermions and bosons is equivalent to ask for the following two equations, one for charged and other for neutral particles:
\begin{eqnarray*}
2 n  &=&  r*s \\
4 n  &=&  r^2+s^2-1
\end{eqnarray*}
which constrains $r=s\pm 1$ and then $n$ must be a triangular number. The obvious extra constraint 
$$n \geq r,s $$
forbids the trivial solution. Thus the minimal solution asks explicitly for three generations, and one and only one
unmatched family, $r=s\pm 1$.

The best scenario should present a \textbf{24} representation decomposing in three [super]multiplets, each with an octet of scalars, and then further symmetry breaking between them.

Lets now to check the ``unoriented string'' sector, ie the diquarks, which must partner with quarks. We can set another two equations again, one for U type and other for D type quarks, respectively DD and UD terminated strings.
\begin{eqnarray*}
 2 n  &=& r(r+1)/2 \\
 2 n &=& r*s
\end{eqnarray*}

The equations are compatible with the previous ones; in this case they ask $2n$ to be in the series of hexagon numbers; the lowest even number in this series is $6$, thus the simplest possibility is $n=3$ as above. 

Furthermore, this second pair brings $2 s = (r+1)$, which jointly with $r=s\pm 1$ \textit{fixes} uniquely
\begin{eqnarray*}
 s &=& 2 \\
 r &=& 3 \\
 n &=& 3
\end{eqnarray*}
which is the exact content of the standard model: three generations with three D type and two U type quarks in the low mass sector.

A problem in this ``unoriented sector'' is that we get three extra diquarks (type UU) with charge $\pm 4/3$, living too in the \textbf{15} representation that comes out from decomposing ${\mathbf5} \otimes {\mathbf 5}$. And one needs some extra argument to get rid of them. For instance, the impossibility of grouping them in a doublet of the $SU(2)$ part of the electroweak group, if it were needed due to the mass mechanism, could be one such argument.

On the positive side, $r+s=5$ is encouraging because the quantization of such string asks for a $SO(2^5)$ group \cite{Marcus:1986cm}, and this $SO(32)$ is the unique group we can use in a theory of Type I superstrings. 

\section{More caveats}

To find a mechanism to give mass to supermultiplets and then at the same time to break them so finely is a serious challenge, because almost every work on SUSY starts from the principle of a hard separation between fermions and their spartners. In this case the neutrinos move away via seesaw, and the electron and the first generation quarks are separated only some few orders of magnitude from the mass eigenstates of the spartner sector, but the muon (and surely its neutrino, before see-saw) is still near of a whole octet of spin 0 particles.

The hint of a SO(32) group raises the doubt about if we must pursue a 4D or a 10D string formalism. Classical superstrings can exist in 4D, but quantum versions ask for 10D. A related theme is the finding by A. Connes and collaborators of a hidden topological dimension in the structure of the fields of the standard model, so that in some aspect they already inhabit a 10 (mod 8) dimensional space. The Maldacena way seems also troublesome: our quark/diquark construction only works with 3 colours, so it is not easy to see how a $N_c\to \infty$ limit, needed to use AdS/CFT dualities, can be built.

In any case, if strings are introduced straightforwardly, one needs a role for the closed string both in the oriented and unoriented sectors. The oriented sector has the peculiarity that an open string can close against itself; such phenomena could explain the link \cite{Rivero:2006rb} between mesons and W and Z particles if we consider that such particles come from the closed strings. And in the same spirit we could hope that the unoriented closed string sector could incorporate the SU(3) glue. But again here there is a departure from traditional approach, because QCD and GSW interactions would be justified from different elements of the theory, instead of a big unified group. Our approach already defies the mass generation at GUT scale; and here also the meaning of the unification of couplings must be reconsiderated. Again, some other approaches to the SM already separate $SU(3)$ color from $SU(2) \otimes U(1)$, on grounds of the vector/chiral character of the interactions.

The limitation of three generations, while a success of the supersymmetry ansatz, strongly constrains the possibilities of an empirical approach; the Rydberg and Balmers of our age can not generate whole series to test their proposals, and we are constrained to play between aesthetic criteria about simplicity and considerations about the order of magnitude of the next perturbation. To my regret, the above section does not amount to a baking recipe for a model. Other sets of coincidences seem to allow for ``4th spectroscopy'', ie preon and other composites managed via relativistic quantum mechanics. But it is hard to extend a model to the whole corpus of data. 

Some minor studies about information entropy have been tried, in order to substitute the human bias of any criterion based on  ``simplicity''. Notably I.J. Good and, more recently, Bailey and Ferguson \cite{BaiFer}. But the approach is not very encouraging: it selects old findings as Lenz's $6\pi^5$ \cite{Lenz} while it is unable to stress the role of groups of relationships as the ones reported here above.

In any case, while it is unwise to base a research project in numerical evidence, it seems also unwise to leave them unreported, at least in preprint format. Readers are invited to browse the unselected, raw collection of data in the above mentioned thread.

\end{document}